\documentclass[aps,prb,twocolumn,superscriptaddress,floatfix]{revtex4-2}
\usepackage[utf8]{inputenc}
\usepackage[english]{babel}
\usepackage{amsmath}
\usepackage{upgreek} 
\usepackage{feynmf}
\usepackage{xcolor}
\usepackage{soul}
\usepackage{graphicx}
\usepackage{amsfonts}
\usepackage{blindtext}
\usepackage{appendix}
\usepackage[colorlinks=true,unicode=true,allcolors=blue]{hyperref}
\usepackage{siunitx}
\usepackage[normalem]{ulem}
\addto\captionsenglish{}

\begin{document}

\preprint{APS/123-QED}
\title{Brillouin-Mandelstam Scattering-based Cooling of Traveling Acoustic Waves from Cryogenic Temperatures}
\author{Lisa Fischer}
 \email[Corresponding author ]{lisa.fischer@mpl.mpg.de}
 \altaffiliation{Authors contributed equally to this work.}
\affiliation{
 Max Planck Institute for the Science of Light, Staudtstr. 2, 91058, Erlangen, Germany}
 \affiliation{Department of Physics, Friedrich-Alexander Universität Erlangen-Nürnberg, Staudtstr. 7, 91058, Erlangen, Germany}
 \affiliation{Institute of Photonics, Leibniz Universität Hannover, Welfengarten 1A 30167, Hannover, Germany}
\author{Laura Blázquez Martínez}
 \email[Corresponding author ]{laura.blazquez@mpl.mpg.de}
 \altaffiliation{Authors contributed equally to this work.}
\affiliation{
 Max Planck Institute for the Science of Light, Staudtstr. 2, 91058, Erlangen, Germany}
 \affiliation{Department of Physics, Friedrich-Alexander Universität Erlangen-Nürnberg, Staudtstr. 7, 91058 Erlangen, Germany}
\affiliation{Institute of Photonics, Leibniz Universität Hannover, Welfengarten 1A 30167, Hannover, Germany}
\author{Robin Chenivière}
\affiliation{Institut des Sciences Chimiques de Rennes, University of Rennes, Campus de Beaulieu, F-35042, Rennes, France}
\author{Johann Troles}
\affiliation{Institut des Sciences Chimiques de Rennes, University of Rennes, Campus de Beaulieu, F-35042, Rennes, France}
\author{Birgit Stiller}
\affiliation{
 Max Planck Institute for the Science of Light, Staudtstr. 2, 91058, Erlangen, Germany}
 \affiliation{Department of Physics, Friedrich-Alexander Universität Erlangen-Nürnberg, Staudtstr. 7, 91058 Erlangen, Germany}
 \affiliation{Institute of Photonics, Leibniz Universität Hannover, Welfengarten 1A 30167, Hannover, Germany}
\date{\today}

\begin{abstract}
Thermal phonons are a major source of decoherence in quantum mechanical systems. Operating in the quantum ground state is therefore often an experimental prerequisite. Additionally to passive cooling in a cryogenic environment, active laser cooling enables the reduction of phonons at specific acoustic frequencies. Brillouin cooling has been used to show efficient reduction of the thermal phonon population in waveguides at GHz frequencies down to 74~K.
In this letter, we demonstrate cooling of a 7.608~GHz acoustic mode by combining Brillouin active cooling with precooling from 77~K using liquid nitrogen. We show a 69\% reduction in the phonon population, resulting in a final temperature of 24.3 ± 1.9\,K, 50\,K lower than previously reported.
\end{abstract}

\maketitle

The coherent transfer of information is a key prerequisite for many quantum technologies such as quantum communication \cite{luo2023recent} ,  quantum storage \cite{wallucks2020quantum, lvovsky2009optical}, detection of gravitational waves \cite{abbott2016gw150914, penrose1996gravity} and ultra precise metrology \cite{teufel2009nanomechanical, mason2019continuous}. To avoid decoherence due to thermal phonons, cooling the experimental platform to the quantum ground state is often needed \cite{o2010quantum, doeleman2023brillouin}. For many applications, passive cooling, such as using cryostats or dilution refrigerators,increases the experimental complexity severely or does not provide sufficient cooling rates, so that additional active cooling techniques are required. \\
To date, most of the research in cooling optomechanical systems has focused on resonators, leading to the demonstration of active cooling of a resonator to its quantum ground state \cite{teufel2011sideband, chan2011laser, qiu2020laser}. However, waveguides are also interesting candidates for quantum technologies, as they have several beneficial properties. Light at any frequency within the transparency window of the optical fiber can be transmitted and is not bound to resonance frequencies. They offer a large bandwidth and low-loss confinement of light over kilometers. Furthermore, they can play both roles: the one of a platform for quantum signal processing and at the same time a transmission line.  Despite these advantages, to the best of our knowledge, active laser cooling in waveguides to temperatures below 74\,K has not yet been demonstrated \cite{blazquez2024optoacoustic}. \\
\begin{figure}[h!]
\centering
\includegraphics[width=\linewidth]{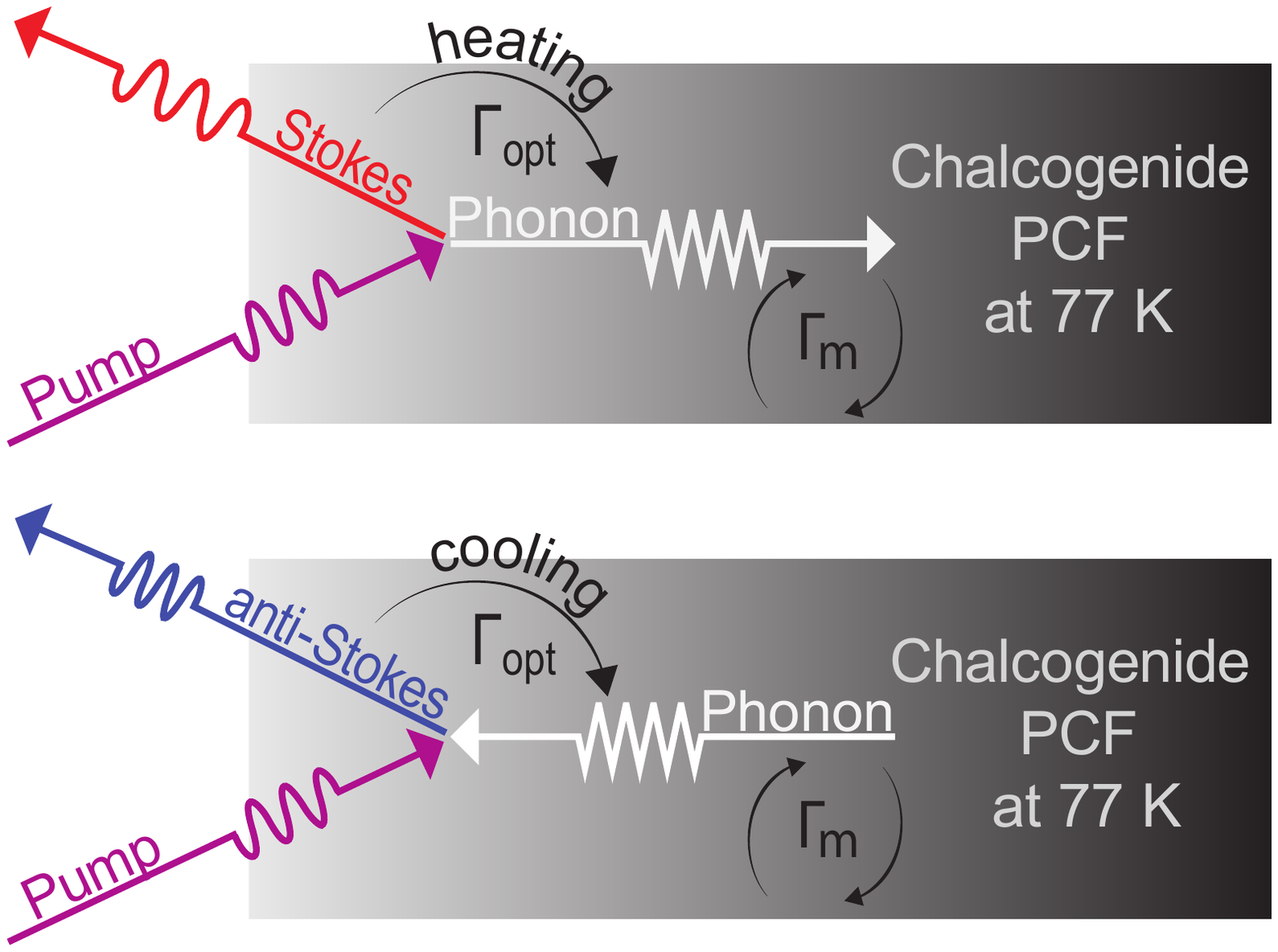}
\caption{Sketch of Stokes (top) and anti-Stokes (bottom)  Brillouin-Mandelstam scattering.}
\label{fig:BrillouinSketch}
\end{figure}

A very promising physical process for actively cooling thermal phonons in waveguides is Brillouin-Mandelstam scattering \cite{zhu2023dynamic,chen2016brillouin}. It is a third-order nonlinear process that coherently couples light and hypersound waves at GHz frequencies \cite{BOYD20081, kobyakov2009stimulated}. The back-scattered spectrum consists of a frequency down-shifted resonance (Stokes) and a frequency up-shifted resonance (anti-Stokes). In anti-Stokes scattering, energy is transferred from the acoustic domain to the optical domain, resulting in cooling of the phonon field. The Stokes interaction, on the other hand, heats the acoustic phonons \cite{WolffTutorial}. A great advantage of Brillouin-based laser cooling in fibers, compared to resonators, is that no techniques to suppress Stokes heating, such as resolved sideband cooling \cite{clark2017sideband, wilson2007theory}, are required to achieve cooling of the acoustic mode. 
This is due to the fact that the Stokes and anti-Stokes processes are decoupled, because they address different phonon modes \cite{zhu2023dynamic}. Stokes scattering heats the phonons copropagating with the pump, while the anti-Stokes process cools the phonons counter-propagating to the pump, as shown in \autoref{fig:BrillouinSketch}. \\
Furthermore, a continuous phonon field is addressed in waveguides, whereby the choice of the corresponding resonant phonons is determined by the pump wavelength \cite{WolffTutorial}. This can be exploited for operations involving multiple wavelengths, such as the storage of multiple optical frequencies \cite{ZhangCohCtrl, stiller2019cross}. Cooling an object that is extended over the whole waveguide also offers the possibility to investigate the transition from classical to quantum physics in macroscopic objects \cite{cattiaux2021macroscopic, MacroscopQuantum}. In addition, waveguides offer full temporal control over the interaction, which makes them very interesting platforms for random access memories \cite{merklein2017chip}.\\
Laser cooling in waveguides due to Brillouin-Mandelstam scattering has already been demonstrated in several experiments starting from room temperature \cite{blazquez2024optoacoustic, Otterstrom, JohnsonCooling}. In \cite{blazquez2024optoacoustic} cooling rates of up to 74.4\% were achieved, reducing the thermal phonon population from 830 to 212. Despite these high cooling rates, the resulting number of phonons is still too large for many quantum experiments.\\
In this work, we demonstrate how lower thermal phonon occupations can be reached by combining active cooling with passive precooling using liquid nitrogen. The fiber used is a tapered photonic crystal fiber (PCF) made out of chalcogenide glass \cite{blazquez2024optoacoustic}. Immersion in liquid nitrogen, which has a temperature of 77\,K, leads to a reduction of the thermal phonon population from 830 phonons at room temperature to 211. Starting from this initial value, we further cooled the acoustic mode by 69\,\% using Brillouin-Mandelstam scattering. This leads to a final phonon number of $66 \pm 6$ phonons. Compared to the phonon number at room temperature, this corresponds to a reduction of 92\%. This provides already an interesting platform for quantum experiments such as optoacoustic entanglement generation \cite{zhu2024optoacoustic}.  

\begin{figure}[t]
    \centering
    \includegraphics[width=
    \linewidth]{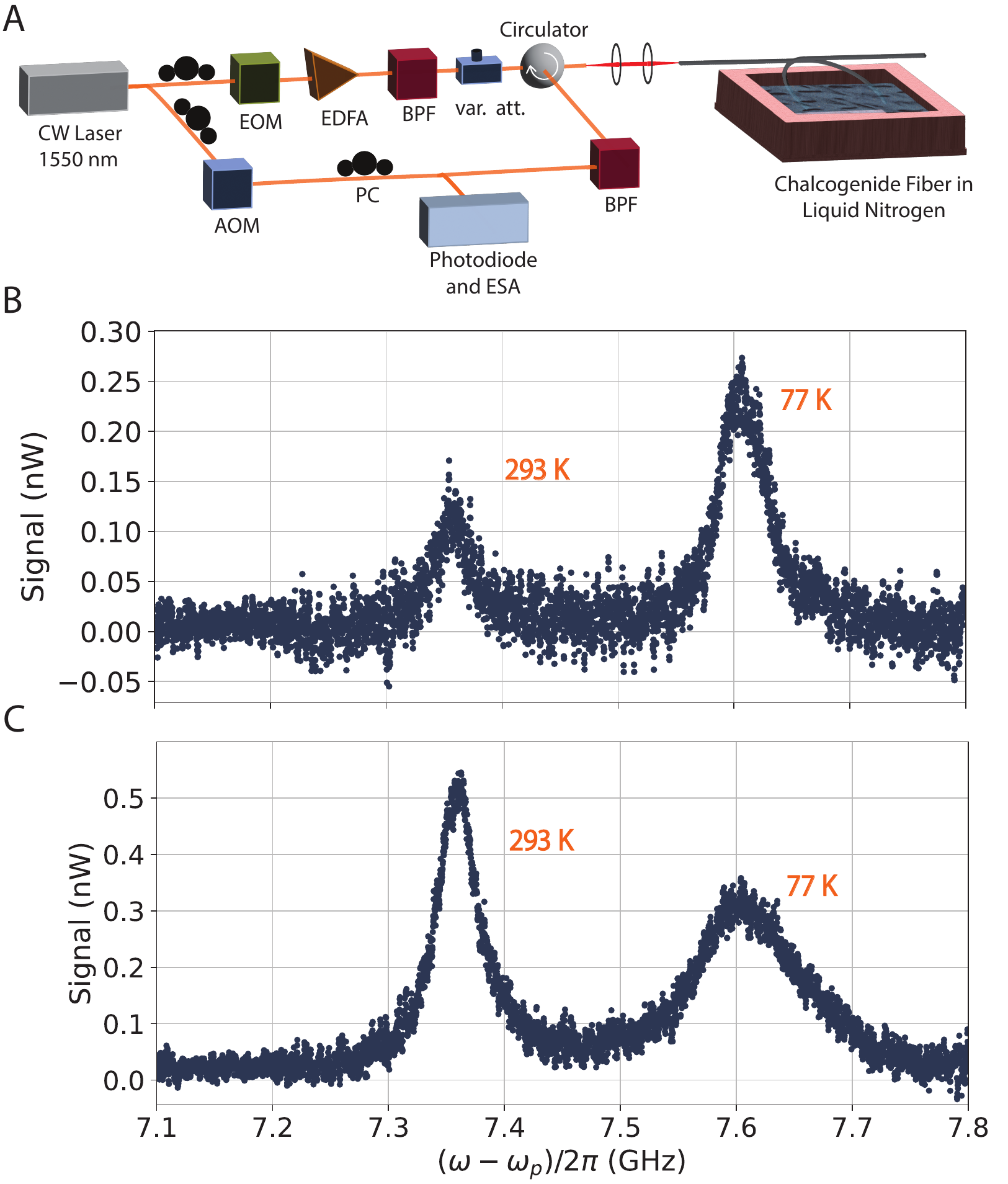}
    \caption{A) Diagram of the experimental Setup. CW: Continuous wave, EOM: electro-optical modulator, EDFA: erbium doped fiber amplifier, BPF: band pass filter, AOM: acousto-optical modulator, PC: polarization controller, ESA: electrical spectrum analyzer, var. att: variable attenuator\\
    B) Back-scattered spectrum of the anti-Stokes process with 9.34\,dBm pump power. The resonance at 7.355$\pm 0.002$\,GHz occurs in the untapered room temperature part of the fiber, whereas the 7.608 $\pm 0.002$\,GHz resonance origins in the tapered part of the fiber at 77\,K.\\
    C) Back-scattered spectrum of the anti-Stokes process at 17.37\,dBm pump power.
    }
    \label{fig:Setup}
\end{figure}

The experimental setup is shown in \autoref{fig:Setup}. A continuous wave (CW) laser emitting at 1550\,nm is divided into two branches, the local oscillator (LO) and the pump branch. The CW pump is modulated into 100\,ns pulses by an electro-optical modulator (EOM). Using long pulses - compared to the length of the fiber - has the advantage of achieving a higher peak power without increasing the average power of the pump and therefore lowering the risk of damaging the surface of the fiber facet. The pulses are amplified by an erbium doped fiber amplifier (EDFA) and filtered by a band pass filter (BPF). The power sent to the fiber is controlled by a variable attenuator and light is launched into the sample via free-space coupling. The sample is a fiber made out of chalcogenide glass with a composition of $\mathrm{Ge_{10}As_{22}Se_{68}}$. To increase the non-linearity within the fiber, its core size is tapered from originally $11\,\mathrm{\mu}$m to a diameter of $2.73\,\mathrm{\mu}$m over a length of $47\pm 2$\,cm. The Brillouin resonance frequency of the fiber is $\Omega_B(2\pi)^{-1} = 7.355 \pm 0.002$\,GHz at room temperature.   
The back-reflected Brillouin signal is separated from the incoming pump light by means of a circulator. After passing through a band-pass filter to filter out the back-reflected pump, the signal is mixed with the LO to perform heterodyne detection. Since the absolute value of the Stokes and anti-Stokes resonance frequency shift from the pump frequency is almost the same and the band-pass filter does not suppress the Stokes resonance completely, the LO is shifted by 200\,MHz to prevent the Stokes and anti-Stokes resonances from overlapping in detection. The beating signal between LO and signal arm is subsequently detected with a photodiode and measured with an electrical spectrum analyzer (ESA). \\
To start the Brillouin cooling from a lower initial thermal phonon population, the fiber is passively pre-cooled. Of all cryogens available, liquid nitrogen stands out, because it is readily available and easy to handle, and is at the low temperature of T = 77\,K. Due to the special geometric structure of PCFs with air holes around the core, the moisture contained in the air freezes in the air holes. This affects the light confinement within the fibers and can lead to high transmission losses.
For the fiber used in our experiment, we were able to demonstrate high transmission even at cryogenic temperatures. The coupling and transmission losses increased by less than 1\,dB from -4.28\,dB at room temperature to -4.69\,dB at 77\,K, when immersing the fiber in liquid nitrogen.\\
Since only the tapered part of the fiber is immersed in liquid nitrogen and the untapered part is at room temperature, the Brillouin spectrum shows two resonances each for Stokes and anti-Stokes measurements. The resonance of the scattering at room temperature is at 7.355\,GHz and 7.608\,GHz at 77\,K. This is shown for the anti-Stokes process in \autoref{fig:Setup}. The resonance at 77K is higher for low powers than the room temperature part of the spectrum. This is due to the fact that the part of the fiber located at 77K is tapered, while the room temperature part is not tapered. The Brillouin gain of the tapered part is significantly higher than that of the untapered part, resulting in a stronger Brillouin response. However, since the number of thermal phonons at 77 K is lower than at room temperature, the resonance at 77 K saturates earlier, resulting in a lower amplitude of the 77 K resonance compared to the resonance at room temperature at higher powers. In the following experiment, only the behavior of the resonance at 77\,K is analyzed.

Without optical perturbation the dissipation rate of the acoustic field is given by the natural acoustic dissipation rate $\Gamma_m$. When an external optical pump is applied, the effective dissipation rate $\Gamma_{eff}$ of the acoustic field is increased by the optical induced losses $\Gamma_{opt}$ caused by the anti-Stokes scattering process $\Gamma_{eff} = \Gamma_{opt} + \Gamma_m$ \cite{blazquez2024optoacoustic}. The reduction of the population $N_b^{ss}$ of the phonon mode cooled by the anti-Stokes scattering can be described mathematically as \cite{blazquez2024optoacoustic}: \\
\begin{equation}
    N_b^{ss} = \frac{4 g_{0m}^2 + \gamma_0 (\Gamma_m + \gamma_0)}{4g_{0m}^2 + \gamma_0\Gamma_m} \cdot \frac{\Gamma_m}{\Gamma_m + \gamma_0} n_{th}\,, \label{eq:theoPhononPopulation}
\end{equation}
with $g_{om}$ the optomechanical coupling strength and $\gamma_0$ the optical dissipation rate.
The cooling rate R of the acoustic mode is defined as \cite{blazquez2024optoacoustic}:
\begin{equation}
    R = \frac{N_b^{ss}}{n_{th}} = \frac{\Gamma_m}{\Gamma_{eff}}\,. \label{eq:CoolingRatio}
\end{equation}
During the experiment, the Brillouin response was measured as a function of optical pump power. The shape of the gain of Stokes and anti-Stokes resonances is given by the Brillouin gain function $G_B(\omega)$ \cite{kobyakov2009stimulated}:\\ 
\begin{equation}
    G_B(\omega) = g_B \frac{\left(\frac{\Gamma_{eff}}{2}\right)^2}{(\Omega_B - \omega)^2 +\left(\frac{\Gamma_{eff}}{2}\right)^2}\,. \label{eq:Gain}
\end{equation}
At low power, Stokes and anti-Stokes resonances grow equally with increasing pump power $\mathrm{P_p}$. At higher power, the Stokes resonance reaches the stimulated regime and grows exponentially. Mathematically, this can be represented as \cite{BOYD20081}:\\
\begin{equation}
    I_S \propto \mathrm{exp}{(\mathrm{G_B I_p L})} \label{eq:expStokesGain}\,,
\end{equation}
with L the length of the fiber, $I_S$ the intensity of the Stokes field and $I_p$ the intensity of the pump field.

\begin{figure}[h]
\centering\includegraphics[width = \linewidth]{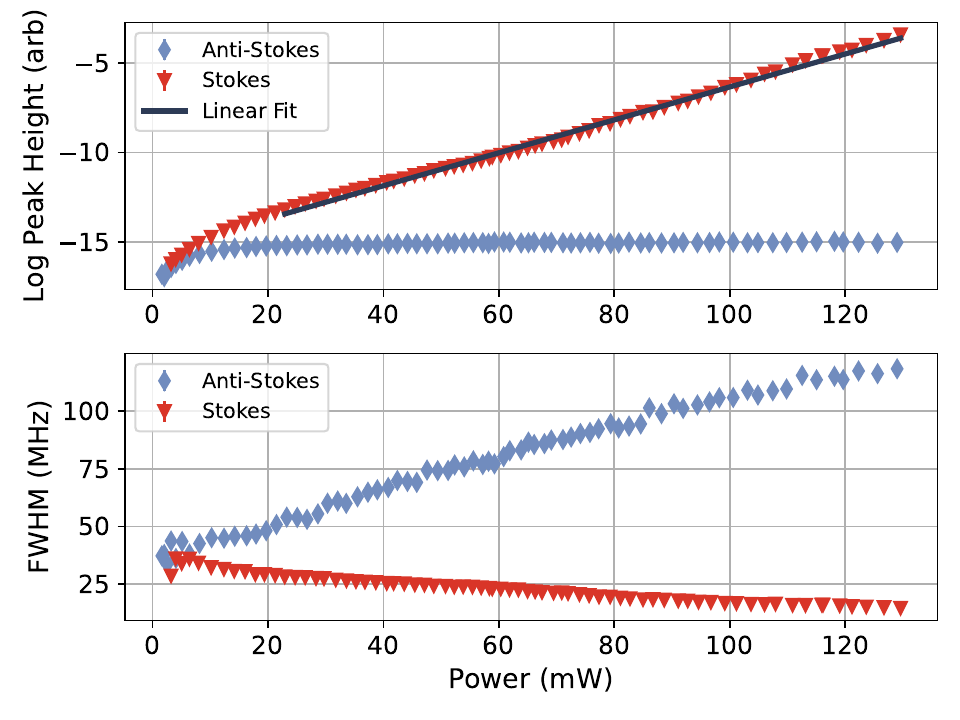}
\caption{Logarithmic peak height (top) and FWHM (bottom) of the Stokes (red triangles) and the anti-Stokes (blue diamonds) resonances as a function of pump power. A linear fit to the linear part of the logarithmic peak height was used to determine the Brillouin gain of the fiber $ G_B = 196.1 \pm 1.9  \,\mathrm{W^{-1}
m^{-1}} $.  }
\label{fig:HeightAndFWHM}
\end{figure}

\begin{figure}
\centering\includegraphics[width = \linewidth]{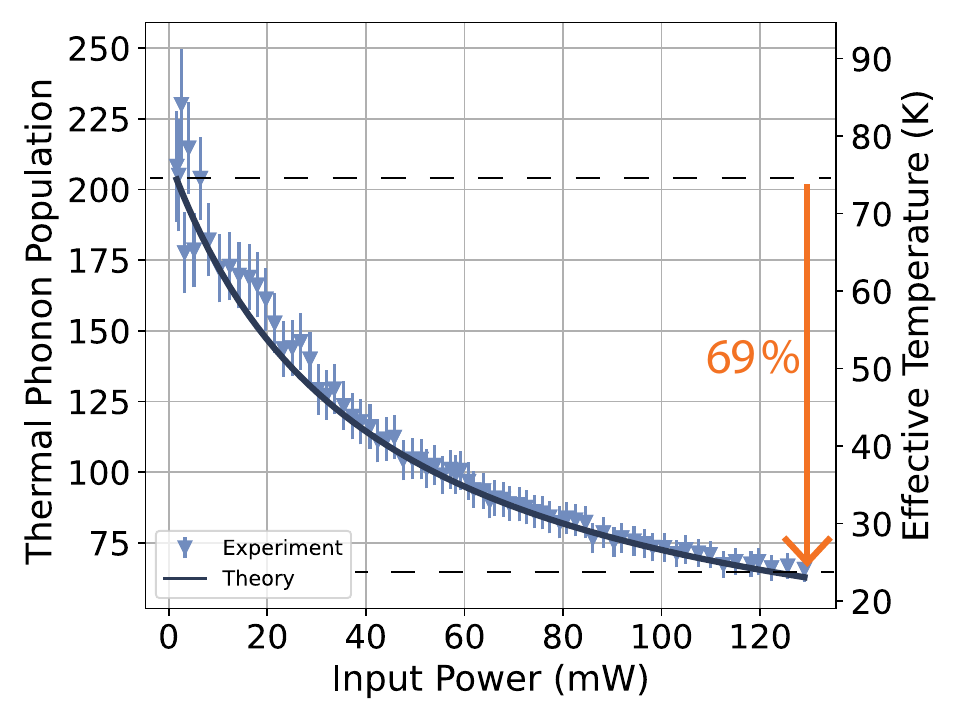}
\caption{Theoretical (solid black line) and measured (blue triangles) thermal phonon population and effective temperature of the $\frac{\Omega_B}{2\pi} = 7.608\,$GHz mode as a function of pump power. At maximum power, the thermal phonon population is reduced by $69\pm3$\,\%, resulting in a final phonon population of $66 \pm 6$ phonons.  }
\label{fig:PhononPopulation}
\end{figure}

Fitting \autoref{eq:Gain} and \autoref{eq:expStokesGain} to the data yields the Brillouin frequency shift $\Omega_B$, the effective dissipation rate $\Gamma_\mathrm{{eff}}$ (FWHM of the resonance) and the peak height as a function of pump power. The FWHM and peak height obtained are plotted versus the pump power in \autoref{fig:HeightAndFWHM}. As expected, Stokes and anti-Stokes initially grow at the same rate as the pumping power increases. Once a certain threshold is reached, the Stokes resonance grows exponentially while the anti-Stokes resonance enters saturation. \\
To calculate the gain of the fiber a linear function is fitted to the exponential part of the peak height of the Stokes resonance. According to \autoref{eq:expStokesGain} its slope m is equal to $\mathrm{m}= \mathrm{G_BL}$. The resulting gain for the chalcogenide PCF at 77\,K is $196.1 \pm 1.9 \,\mathrm{W^{-1} m^{-1}}$. This is slightly higher than the gain measured at room temperature $G_B = 178.4 \pm 1.6 \,\mathrm{W^{-1}
m^{-1}}$. This corresponds to a Brillouin-Mandelstam material gain of $g_B = (1.044 \pm 0.010)\cdot 10^{-9} \,\mathrm{W^{-1}} \mathrm{m}$ at room temperature, comparable to literature values \cite{tow2012linewidth, tow2013laser}. \\
The natural acoustic dissipation rate determined from \autoref{fig:HeightAndFWHM} is $36.8\pm 2.5$\,MHz. This is narrower than the 49.3 $\pm$ 0.2\,MHz measured at room temperature. 
At room temperature the Brillouin-Mandelstam resonance of the fiber is centered at $7.355 \pm 0.002$\,GHz. When placed in a bath with liquid nitrogen, it shifts to 7.608 $\pm$ 0.002\,GHz. The dependence of the Brillouin frequency shift and linewidth on temperature has already been observed in silica \cite{cryer2025brillouin, le2001experimental, le2003study}. In the chalcogenide PCF used in this experiment, the Brillouin frequency shifts to higher frequencies with decreasing temperature, instead of lower frequencies as shown in silica. A similar behavior has already been observed for coated chalcogenide fibers at room temperature \cite{wang2020stimulated}. \\
The average number of phonons $n_{th}$ populating the 7.608\,GHz acoustic mode at temperature T = 77\,K can be calculated using Bose-Einstein-statistics \cite{WolffTutorial}:\\
\begin{equation}
    n_{th} = \frac{1}{\mathrm{exp}( {\frac{\hbar \Omega}{k_B T}})-1}\,, \label{eq:BoseEinstein}
\end{equation}
with $\hbar$ the Planck constant and $k_B$ the Boltzmann constant.
The resulting initial phonon occupation of the 77\,K resonance is 211 phonons. 
Using this together with \autoref{eq:CoolingRatio} and the measured $\Gamma_{eff}$, the resulting phonon population as a function of pump power and the corresponding effective temperature can be calculated. This is shown in \autoref{fig:PhononPopulation}. With increasing pump power the acoustic mode cools down, until reaching an average population of $66 \pm 6$ phonons. This corresponds to a cooling rate of $69\pm3\,$ \% of the 77\,K mode, resulting in a final effective temperature of $24.3 \pm 1.9$\,K. The theory curve was calculated using \autoref{eq:theoPhononPopulation}, in good agreement with the experiment. 

In this letter, we have shown that when combining Brillouin-Mandelstam scattering with a liquid nitrogen environment, low phonon populations can be reached. This is the first demonstration of active Brillouin cooling in linear waveguides operating in a cryogenic environment. The sample used was a PCF chalcogenide with a glass composition of $\mathrm{Ge_{10}As_{22}Se_{68}}$, tapered to a core diameter of 2.73\,$\mu$m over a length of 47\,cm and passively cooled to 77\,K using liquid nitrogen. It was shown that, starting from this temperature, an acoustic mode can be cooled down to $24.3 \pm 1.9$\,K by means of active Brillouin cooling. This corresponds to a reduction of the thermal phonon population of this mode of $69\pm3$\,\% from the original 211 phonons to $66 \pm 6$. This provides a potential platform for quantum experiments like optoacoustic entanglement \cite{zhu2024optoacoustic}. The high Brillouin gain of the fiber at 77\,K makes it furthermore interesting for optoacoustic experiments in the strong coupling regime \cite{martínez2025cavitylessbrillouinstrongcoupling}. To achieve even higher cooling rates down to single phonons, Brillouin-Mandelstam scattering could be combined with passive cooling at lower cryogenic temperatures. This will open up new research opportunities in the field of fiber-based quantum applications such as entanglement generation between phonons \cite{chou2025phononEntanglement}, quantum transducer \cite{mirhosseini2020superconducting} and quantum memories \cite{wallucks2020quantum, guo2019highQuantMemory}. \\

\begin{acknowledgments}

We thank C. Zhu, X. Zeng and G. Slinkov for their help at different stages of
the experiment. We thank J. H. Marines Cabello for his participation in the quest of obtaining the sample. \\
The authors declare no conflicts of interest. \\
Data underlying the results presented in this paper are not publicly available at this time but may be obtained from the authors upon reasonable request.

\end{acknowledgments}


\begin{thebibliography}{38}%
\makeatletter
\providecommand \@ifxundefined [1]{%
 \@ifx{#1\undefined}
}%
\providecommand \@ifnum [1]{%
 \ifnum #1\expandafter \@firstoftwo
 \else \expandafter \@secondoftwo
 \fi
}%
\providecommand \@ifx [1]{%
 \ifx #1\expandafter \@firstoftwo
 \else \expandafter \@secondoftwo
 \fi
}%
\providecommand \natexlab [1]{#1}%
\providecommand \enquote  [1]{``#1''}%
\providecommand \bibnamefont  [1]{#1}%
\providecommand \bibfnamefont [1]{#1}%
\providecommand \citenamefont [1]{#1}%
\providecommand \href@noop [0]{\@secondoftwo}%
\providecommand \href [0]{\begingroup \@sanitize@url \@href}%
\providecommand \@href[1]{\@@startlink{#1}\@@href}%
\providecommand \@@href[1]{\endgroup#1\@@endlink}%
\providecommand \@sanitize@url [0]{\catcode `\\12\catcode `\$12\catcode
  `\&12\catcode `\#12\catcode `\^12\catcode `\_12\catcode `\%12\relax}%
\providecommand \@@startlink[1]{}%
\providecommand \@@endlink[0]{}%
\providecommand \url  [0]{\begingroup\@sanitize@url \@url }%
\providecommand \@url [1]{\endgroup\@href {#1}{\urlprefix }}%
\providecommand \urlprefix  [0]{URL }%
\providecommand \Eprint [0]{\href }%
\providecommand \doibase [0]{https://doi.org/}%
\providecommand \selectlanguage [0]{\@gobble}%
\providecommand \bibinfo  [0]{\@secondoftwo}%
\providecommand \bibfield  [0]{\@secondoftwo}%
\providecommand \translation [1]{[#1]}%
\providecommand \BibitemOpen [0]{}%
\providecommand \bibitemStop [0]{}%
\providecommand \bibitemNoStop [0]{.\EOS\space}%
\providecommand \EOS [0]{\spacefactor3000\relax}%
\providecommand \BibitemShut  [1]{\csname bibitem#1\endcsname}%
\let\auto@bib@innerbib\@empty
\bibitem [{\citenamefont {Luo}\ \emph {et~al.}(2023)\citenamefont {Luo},
  \citenamefont {Cao}, \citenamefont {Shi}, \citenamefont {Wan}, \citenamefont
  {Zhang}, \citenamefont {Li}, \citenamefont {Chen}, \citenamefont {Li},
  \citenamefont {Li}, \citenamefont {Wang}, \citenamefont {Sun}, \citenamefont
  {Karim}, \citenamefont {Cai}, \citenamefont {Kwek},\ and\ \citenamefont
  {Liu}}]{luo2023recent}%
  \BibitemOpen
  \bibfield  {author} {\bibinfo {author} {\bibfnamefont {W.}~\bibnamefont
  {Luo}}, \bibinfo {author} {\bibfnamefont {L.}~\bibnamefont {Cao}}, \bibinfo
  {author} {\bibfnamefont {Y.}~\bibnamefont {Shi}}, \bibinfo {author}
  {\bibfnamefont {L.}~\bibnamefont {Wan}}, \bibinfo {author} {\bibfnamefont
  {H.}~\bibnamefont {Zhang}}, \bibinfo {author} {\bibfnamefont
  {S.}~\bibnamefont {Li}}, \bibinfo {author} {\bibfnamefont {G.}~\bibnamefont
  {Chen}}, \bibinfo {author} {\bibfnamefont {Y.}~\bibnamefont {Li}}, \bibinfo
  {author} {\bibfnamefont {S.}~\bibnamefont {Li}}, \bibinfo {author}
  {\bibfnamefont {Y.}~\bibnamefont {Wang}}, \bibinfo {author} {\bibfnamefont
  {S.}~\bibnamefont {Sun}}, \bibinfo {author} {\bibfnamefont {M.~F.}\
  \bibnamefont {Karim}}, \bibinfo {author} {\bibfnamefont {H.}~\bibnamefont
  {Cai}}, \bibinfo {author} {\bibfnamefont {L.~C.}\ \bibnamefont {Kwek}},\ and\
  \bibinfo {author} {\bibfnamefont {A.~Q.}\ \bibnamefont {Liu}},\ }\bibfield
  {title} {\bibinfo {title} {Recent progress in quantum photonic chips for
  quantum communication and internet},\ }\href
  {https://doi.org/10.1038/s41377-023-01173-8} {\bibfield  {journal} {\bibinfo
  {journal} {Light: Science \& Applications}\ }\textbf {\bibinfo {volume}
  {12}},\ \bibinfo {pages} {175} (\bibinfo {year} {2023})}\BibitemShut
  {NoStop}%
\bibitem [{\citenamefont {Wallucks}\ \emph {et~al.}(2020)\citenamefont
  {Wallucks}, \citenamefont {Marinkovi{\'c}}, \citenamefont {Hensen},
  \citenamefont {Stockill},\ and\ \citenamefont
  {Gr{\"o}blacher}}]{wallucks2020quantum}%
  \BibitemOpen
  \bibfield  {author} {\bibinfo {author} {\bibfnamefont {A.}~\bibnamefont
  {Wallucks}}, \bibinfo {author} {\bibfnamefont {I.}~\bibnamefont
  {Marinkovi{\'c}}}, \bibinfo {author} {\bibfnamefont {B.}~\bibnamefont
  {Hensen}}, \bibinfo {author} {\bibfnamefont {R.}~\bibnamefont {Stockill}},\
  and\ \bibinfo {author} {\bibfnamefont {S.}~\bibnamefont {Gr{\"o}blacher}},\
  }\bibfield  {title} {\bibinfo {title} {A quantum memory at telecom
  wavelengths},\ }\href {https://doi.org/10.1038/s41567-020-0891-z} {\bibfield
  {journal} {\bibinfo  {journal} {Nature Physics}\ }\textbf {\bibinfo {volume}
  {16}},\ \bibinfo {pages} {772} (\bibinfo {year} {2020})}\BibitemShut
  {NoStop}%
\bibitem [{\citenamefont {Lvovsky}\ \emph {et~al.}(2009)\citenamefont
  {Lvovsky}, \citenamefont {Sanders},\ and\ \citenamefont
  {Tittel}}]{lvovsky2009optical}%
  \BibitemOpen
  \bibfield  {author} {\bibinfo {author} {\bibfnamefont {A.~I.}\ \bibnamefont
  {Lvovsky}}, \bibinfo {author} {\bibfnamefont {B.~C.}\ \bibnamefont
  {Sanders}},\ and\ \bibinfo {author} {\bibfnamefont {W.}~\bibnamefont
  {Tittel}},\ }\bibfield  {title} {\bibinfo {title} {Optical quantum memory},\
  }\href {https://doi.org/10.1038/nphoton.2009.231} {\bibfield  {journal}
  {\bibinfo  {journal} {Nature Photonics}\ }\textbf {\bibinfo {volume} {3}},\
  \bibinfo {pages} {706} (\bibinfo {year} {2009})}\BibitemShut {NoStop}%
\bibitem [{\citenamefont {Abbott}\ \emph {et~al.}(2016)\citenamefont {Abbott},
  \citenamefont {Abbott}, \citenamefont {Abbott}, \citenamefont {Abernathy},
  \citenamefont {Acernese}, \citenamefont {Ackley}, \citenamefont {Adams},
  \citenamefont {Adams}, \citenamefont {Addesso}, \citenamefont {Adhikari}
  \emph {et~al.}}]{abbott2016gw150914}%
  \BibitemOpen
  \bibfield  {author} {\bibinfo {author} {\bibfnamefont {B.~P.}\ \bibnamefont
  {Abbott}}, \bibinfo {author} {\bibfnamefont {R.}~\bibnamefont {Abbott}},
  \bibinfo {author} {\bibfnamefont {{\relax TD}.}~\bibnamefont {Abbott}},
  \bibinfo {author} {\bibfnamefont {{\relax MR}.}~\bibnamefont {Abernathy}},
  \bibinfo {author} {\bibfnamefont {F.}~\bibnamefont {Acernese}}, \bibinfo
  {author} {\bibfnamefont {K.}~\bibnamefont {Ackley}}, \bibinfo {author}
  {\bibfnamefont {C.}~\bibnamefont {Adams}}, \bibinfo {author} {\bibfnamefont
  {T.}~\bibnamefont {Adams}}, \bibinfo {author} {\bibfnamefont
  {P.}~\bibnamefont {Addesso}}, \bibinfo {author} {\bibfnamefont {{\relax
  RX}.}~\bibnamefont {Adhikari}}, \emph {et~al.},\ }\bibfield  {title}
  {\bibinfo {title} {{{GW150914}}: {{First}} results from the search for binary
  black hole coalescence with {{Advanced LIGO}}},\ }\href
  {https://doi.org/10.1103/PhysRevD.93.122003} {\bibfield  {journal} {\bibinfo
  {journal} {Physical Review D}\ }\textbf {\bibinfo {volume} {93}},\ \bibinfo
  {pages} {122003} (\bibinfo {year} {2016})}\BibitemShut {NoStop}%
\bibitem [{\citenamefont {Penrose}(1996)}]{penrose1996gravity}%
  \BibitemOpen
  \bibfield  {author} {\bibinfo {author} {\bibfnamefont {R.}~\bibnamefont
  {Penrose}},\ }\bibfield  {title} {\bibinfo {title} {On {{Gravity}}'s role in
  {{Quantum State Reduction}}},\ }\href {https://doi.org/10.1007/BF02105068}
  {\bibfield  {journal} {\bibinfo  {journal} {General Relativity and
  Gravitation}\ }\textbf {\bibinfo {volume} {28}},\ \bibinfo {pages} {581}
  (\bibinfo {year} {1996})}\BibitemShut {NoStop}%
\bibitem [{\citenamefont {Teufel}\ \emph {et~al.}(2009)\citenamefont {Teufel},
  \citenamefont {Donner}, \citenamefont {{Castellanos-Beltran}}, \citenamefont
  {Harlow},\ and\ \citenamefont {Lehnert}}]{teufel2009nanomechanical}%
  \BibitemOpen
  \bibfield  {author} {\bibinfo {author} {\bibfnamefont {J.~D.}\ \bibnamefont
  {Teufel}}, \bibinfo {author} {\bibfnamefont {T.}~\bibnamefont {Donner}},
  \bibinfo {author} {\bibfnamefont {M.~A.}\ \bibnamefont
  {{Castellanos-Beltran}}}, \bibinfo {author} {\bibfnamefont {J.~W.}\
  \bibnamefont {Harlow}},\ and\ \bibinfo {author} {\bibfnamefont {K.~W.}\
  \bibnamefont {Lehnert}},\ }\bibfield  {title} {\bibinfo {title}
  {Nanomechanical motion measured with an imprecision below that at the
  standard quantum limit},\ }\href {https://doi.org/10.1038/nnano.2009.343}
  {\bibfield  {journal} {\bibinfo  {journal} {Nature Nanotechnology}\ }\textbf
  {\bibinfo {volume} {4}},\ \bibinfo {pages} {820} (\bibinfo {year}
  {2009})}\BibitemShut {NoStop}%
\bibitem [{\citenamefont {Mason}\ \emph {et~al.}(2019)\citenamefont {Mason},
  \citenamefont {Chen}, \citenamefont {Rossi}, \citenamefont {Tsaturyan},\ and\
  \citenamefont {Schliesser}}]{mason2019continuous}%
  \BibitemOpen
  \bibfield  {author} {\bibinfo {author} {\bibfnamefont {D.}~\bibnamefont
  {Mason}}, \bibinfo {author} {\bibfnamefont {J.}~\bibnamefont {Chen}},
  \bibinfo {author} {\bibfnamefont {M.}~\bibnamefont {Rossi}}, \bibinfo
  {author} {\bibfnamefont {Y.}~\bibnamefont {Tsaturyan}},\ and\ \bibinfo
  {author} {\bibfnamefont {A.}~\bibnamefont {Schliesser}},\ }\bibfield  {title}
  {\bibinfo {title} {Continuous force and displacement measurement below the
  standard quantum limit},\ }\href {https://doi.org/10.1038/s41567-019-0533-5}
  {\bibfield  {journal} {\bibinfo  {journal} {Nature Physics}\ }\textbf
  {\bibinfo {volume} {15}},\ \bibinfo {pages} {745} (\bibinfo {year}
  {2019})}\BibitemShut {NoStop}%
\bibitem [{\citenamefont {O'Connell}\ \emph {et~al.}(2010)\citenamefont
  {O'Connell}, \citenamefont {Hofheinz}, \citenamefont {Ansmann}, \citenamefont
  {Bialczak}, \citenamefont {Lenander}, \citenamefont {Lucero}, \citenamefont
  {Neeley}, \citenamefont {Sank}, \citenamefont {Wang}, \citenamefont {Weides},
  \citenamefont {Wenner}, \citenamefont {Martinis},\ and\ \citenamefont
  {Cleland}}]{o2010quantum}%
  \BibitemOpen
  \bibfield  {author} {\bibinfo {author} {\bibfnamefont {A.~D.}\ \bibnamefont
  {O'Connell}}, \bibinfo {author} {\bibfnamefont {M.}~\bibnamefont {Hofheinz}},
  \bibinfo {author} {\bibfnamefont {M.}~\bibnamefont {Ansmann}}, \bibinfo
  {author} {\bibfnamefont {R.~C.}\ \bibnamefont {Bialczak}}, \bibinfo {author}
  {\bibfnamefont {M.}~\bibnamefont {Lenander}}, \bibinfo {author}
  {\bibfnamefont {E.}~\bibnamefont {Lucero}}, \bibinfo {author} {\bibfnamefont
  {M.}~\bibnamefont {Neeley}}, \bibinfo {author} {\bibfnamefont
  {D.}~\bibnamefont {Sank}}, \bibinfo {author} {\bibfnamefont {H.}~\bibnamefont
  {Wang}}, \bibinfo {author} {\bibfnamefont {M.}~\bibnamefont {Weides}},
  \bibinfo {author} {\bibfnamefont {J.}~\bibnamefont {Wenner}}, \bibinfo
  {author} {\bibfnamefont {J.~M.}\ \bibnamefont {Martinis}},\ and\ \bibinfo
  {author} {\bibfnamefont {A.~N.}\ \bibnamefont {Cleland}},\ }\bibfield
  {title} {\bibinfo {title} {Quantum ground state and single-phonon control of
  a mechanical resonator},\ }\href {https://doi.org/10.1038/nature08967}
  {\bibfield  {journal} {\bibinfo  {journal} {Nature}\ }\textbf {\bibinfo
  {volume} {464}},\ \bibinfo {pages} {697} (\bibinfo {year}
  {2010})}\BibitemShut {NoStop}%
\bibitem [{\citenamefont {Doeleman}\ \emph {et~al.}(2023)\citenamefont
  {Doeleman}, \citenamefont {Schatteburg}, \citenamefont {Benevides},
  \citenamefont {Vollenweider}, \citenamefont {Macri},\ and\ \citenamefont
  {Chu}}]{doeleman2023brillouin}%
  \BibitemOpen
  \bibfield  {author} {\bibinfo {author} {\bibfnamefont {H.~M.}\ \bibnamefont
  {Doeleman}}, \bibinfo {author} {\bibfnamefont {T.}~\bibnamefont
  {Schatteburg}}, \bibinfo {author} {\bibfnamefont {R.}~\bibnamefont
  {Benevides}}, \bibinfo {author} {\bibfnamefont {S.}~\bibnamefont
  {Vollenweider}}, \bibinfo {author} {\bibfnamefont {D.}~\bibnamefont
  {Macri}},\ and\ \bibinfo {author} {\bibfnamefont {Y.}~\bibnamefont {Chu}},\
  }\bibfield  {title} {\bibinfo {title} {Brillouin optomechanics in the quantum
  ground state},\ }\href {https://doi.org/10.1103/PhysRevResearch.5.043140}
  {\bibfield  {journal} {\bibinfo  {journal} {Physical Review Research}\
  }\textbf {\bibinfo {volume} {5}},\ \bibinfo {pages} {043140} (\bibinfo {year}
  {2023})}\BibitemShut {NoStop}%
\bibitem [{\citenamefont {Teufel}\ \emph {et~al.}(2011)\citenamefont {Teufel},
  \citenamefont {Donner}, \citenamefont {Li}, \citenamefont {Harlow},
  \citenamefont {Allman}, \citenamefont {Cicak}, \citenamefont {Sirois},
  \citenamefont {Whittaker}, \citenamefont {Lehnert},\ and\ \citenamefont
  {Simmonds}}]{teufel2011sideband}%
  \BibitemOpen
  \bibfield  {author} {\bibinfo {author} {\bibfnamefont {J.~D.}\ \bibnamefont
  {Teufel}}, \bibinfo {author} {\bibfnamefont {T.}~\bibnamefont {Donner}},
  \bibinfo {author} {\bibfnamefont {D.}~\bibnamefont {Li}}, \bibinfo {author}
  {\bibfnamefont {J.~W.}\ \bibnamefont {Harlow}}, \bibinfo {author}
  {\bibfnamefont {M.~S.}\ \bibnamefont {Allman}}, \bibinfo {author}
  {\bibfnamefont {K.}~\bibnamefont {Cicak}}, \bibinfo {author} {\bibfnamefont
  {A.~J.}\ \bibnamefont {Sirois}}, \bibinfo {author} {\bibfnamefont {J.~D.}\
  \bibnamefont {Whittaker}}, \bibinfo {author} {\bibfnamefont {K.~W.}\
  \bibnamefont {Lehnert}},\ and\ \bibinfo {author} {\bibfnamefont {R.~W.}\
  \bibnamefont {Simmonds}},\ }\bibfield  {title} {\bibinfo {title} {Sideband
  cooling of micromechanical motion to the quantum ground state},\ }\href
  {https://doi.org/10.1038/nature10261} {\bibfield  {journal} {\bibinfo
  {journal} {Nature}\ }\textbf {\bibinfo {volume} {475}},\ \bibinfo {pages}
  {359} (\bibinfo {year} {2011})}\BibitemShut {NoStop}%
\bibitem [{\citenamefont {Chan}\ \emph {et~al.}(2011)\citenamefont {Chan},
  \citenamefont {Alegre}, \citenamefont {{Safavi-Naeini}}, \citenamefont
  {Hill}, \citenamefont {Krause}, \citenamefont {Gr{\"o}blacher}, \citenamefont
  {Aspelmeyer},\ and\ \citenamefont {Painter}}]{chan2011laser}%
  \BibitemOpen
  \bibfield  {author} {\bibinfo {author} {\bibfnamefont {J.}~\bibnamefont
  {Chan}}, \bibinfo {author} {\bibfnamefont {T.~P.~M.}\ \bibnamefont {Alegre}},
  \bibinfo {author} {\bibfnamefont {A.~H.}\ \bibnamefont {{Safavi-Naeini}}},
  \bibinfo {author} {\bibfnamefont {J.~T.}\ \bibnamefont {Hill}}, \bibinfo
  {author} {\bibfnamefont {A.}~\bibnamefont {Krause}}, \bibinfo {author}
  {\bibfnamefont {S.}~\bibnamefont {Gr{\"o}blacher}}, \bibinfo {author}
  {\bibfnamefont {M.}~\bibnamefont {Aspelmeyer}},\ and\ \bibinfo {author}
  {\bibfnamefont {O.}~\bibnamefont {Painter}},\ }\bibfield  {title} {\bibinfo
  {title} {Laser cooling of a nanomechanical oscillator into its quantum ground
  state},\ }\href {https://doi.org/10.1038/nature10461} {\bibfield  {journal}
  {\bibinfo  {journal} {Nature}\ }\textbf {\bibinfo {volume} {478}},\ \bibinfo
  {pages} {89} (\bibinfo {year} {2011})}\BibitemShut {NoStop}%
\bibitem [{\citenamefont {Qiu}\ \emph {et~al.}(2020)\citenamefont {Qiu},
  \citenamefont {Shomroni}, \citenamefont {Seidler},\ and\ \citenamefont
  {Kippenberg}}]{qiu2020laser}%
  \BibitemOpen
  \bibfield  {author} {\bibinfo {author} {\bibfnamefont {L.}~\bibnamefont
  {Qiu}}, \bibinfo {author} {\bibfnamefont {I.}~\bibnamefont {Shomroni}},
  \bibinfo {author} {\bibfnamefont {P.}~\bibnamefont {Seidler}},\ and\ \bibinfo
  {author} {\bibfnamefont {T.~J.}\ \bibnamefont {Kippenberg}},\ }\bibfield
  {title} {\bibinfo {title} {Laser {{Cooling}} of a {{Nanomechanical
  Oscillator}} to {{Its Zero-Point Energy}}},\ }\href
  {https://doi.org/10.1103/PhysRevLett.124.173601} {\bibfield  {journal}
  {\bibinfo  {journal} {Physical Review Letters}\ }\textbf {\bibinfo {volume}
  {124}},\ \bibinfo {pages} {173601} (\bibinfo {year} {2020})}\BibitemShut
  {NoStop}%
\bibitem [{\citenamefont {Bl{\'a}zquez~Mart{\'i}nez}\ \emph
  {et~al.}(2024)\citenamefont {Bl{\'a}zquez~Mart{\'i}nez}, \citenamefont
  {Wiedemann}, \citenamefont {Zhu}, \citenamefont {Geilen},\ and\ \citenamefont
  {Stiller}}]{blazquez2024optoacoustic}%
  \BibitemOpen
  \bibfield  {author} {\bibinfo {author} {\bibfnamefont {L.}~\bibnamefont
  {Bl{\'a}zquez~Mart{\'i}nez}}, \bibinfo {author} {\bibfnamefont
  {P.}~\bibnamefont {Wiedemann}}, \bibinfo {author} {\bibfnamefont
  {C.}~\bibnamefont {Zhu}}, \bibinfo {author} {\bibfnamefont {A.}~\bibnamefont
  {Geilen}},\ and\ \bibinfo {author} {\bibfnamefont {B.}~\bibnamefont
  {Stiller}},\ }\bibfield  {title} {\bibinfo {title} {Optoacoustic {{Cooling}}
  of {{Traveling Hypersound Waves}}},\ }\href
  {https://doi.org/10.1103/PhysRevLett.132.023603} {\bibfield  {journal}
  {\bibinfo  {journal} {Physical Review Letters}\ }\textbf {\bibinfo {volume}
  {132}},\ \bibinfo {pages} {023603} (\bibinfo {year} {2024})}\BibitemShut
  {NoStop}%
\bibitem [{\citenamefont {Zhu}\ and\ \citenamefont
  {Stiller}(2023)}]{zhu2023dynamic}%
  \BibitemOpen
  \bibfield  {author} {\bibinfo {author} {\bibfnamefont {C.}~\bibnamefont
  {Zhu}}\ and\ \bibinfo {author} {\bibfnamefont {B.}~\bibnamefont {Stiller}},\
  }\bibfield  {title} {\bibinfo {title} {Dynamic {{Brillouin}} cooling for
  continuous optomechanical systems},\ }\href
  {https://doi.org/10.1088/2633-4356/acc2a5} {\bibfield  {journal} {\bibinfo
  {journal} {Materials for Quantum Technology}\ }\textbf {\bibinfo {volume}
  {3}},\ \bibinfo {pages} {015003} (\bibinfo {year} {2023})}\BibitemShut
  {NoStop}%
\bibitem [{\citenamefont {Chen}\ \emph {et~al.}(2016)\citenamefont {Chen},
  \citenamefont {Kim},\ and\ \citenamefont {Bahl}}]{chen2016brillouin}%
  \BibitemOpen
  \bibfield  {author} {\bibinfo {author} {\bibfnamefont {Y.-C.}\ \bibnamefont
  {Chen}}, \bibinfo {author} {\bibfnamefont {S.}~\bibnamefont {Kim}},\ and\
  \bibinfo {author} {\bibfnamefont {G.}~\bibnamefont {Bahl}},\ }\bibfield
  {title} {\bibinfo {title} {Brillouin cooling in a linear waveguide},\ }\href
  {https://doi.org/10.1088/1367-2630/18/11/115004} {\bibfield  {journal}
  {\bibinfo  {journal} {New Journal of Physics}\ }\textbf {\bibinfo {volume}
  {18}},\ \bibinfo {pages} {115004} (\bibinfo {year} {2016})}\BibitemShut
  {NoStop}%
\bibitem [{\citenamefont {Boyd}\ \emph {et~al.}(2023)\citenamefont {Boyd},
  \citenamefont {Gaeta},\ and\ \citenamefont {Giese}}]{BOYD20081}%
  \BibitemOpen
  \bibfield  {author} {\bibinfo {author} {\bibfnamefont {R.~W.}\ \bibnamefont
  {Boyd}}, \bibinfo {author} {\bibfnamefont {A.~L.}\ \bibnamefont {Gaeta}},\
  and\ \bibinfo {author} {\bibfnamefont {E.}~\bibnamefont {Giese}},\ }\bibfield
   {title} {\bibinfo {title} {Nonlinear {{Optics}}},\ }in\ \href
  {https://doi.org/10.1007/978-3-030-73893-8_76} {\emph {\bibinfo {booktitle}
  {Springer {{Handbook}} of {{Atomic}}, {{Molecular}}, and {{Optical
  Physics}}}}},\ \bibinfo {editor} {edited by\ \bibinfo {editor} {\bibfnamefont
  {G.~W.~F.}\ \bibnamefont {Drake}}}\ (\bibinfo  {publisher} {Springer
  International Publishing},\ \bibinfo {address} {Cham},\ \bibinfo {year}
  {2023})\ pp.\ \bibinfo {pages} {1097--1110}\BibitemShut {NoStop}%
\bibitem [{\citenamefont {Kobyakov}\ \emph {et~al.}(2010)\citenamefont
  {Kobyakov}, \citenamefont {Sauer},\ and\ \citenamefont
  {Chowdhury}}]{kobyakov2009stimulated}%
  \BibitemOpen
  \bibfield  {author} {\bibinfo {author} {\bibfnamefont {A.}~\bibnamefont
  {Kobyakov}}, \bibinfo {author} {\bibfnamefont {M.}~\bibnamefont {Sauer}},\
  and\ \bibinfo {author} {\bibfnamefont {D.}~\bibnamefont {Chowdhury}},\
  }\bibfield  {title} {\bibinfo {title} {Stimulated {{Brillouin}} scattering in
  optical fibers},\ }\href {https://doi.org/10.1364/AOP.2.000001} {\bibfield
  {journal} {\bibinfo  {journal} {Advances in Optics and Photonics}\ }\textbf
  {\bibinfo {volume} {2}},\ \bibinfo {pages} {1} (\bibinfo {year}
  {2010})}\BibitemShut {NoStop}%
\bibitem [{\citenamefont {Wolff}\ \emph {et~al.}(2021)\citenamefont {Wolff},
  \citenamefont {Smith}, \citenamefont {Stiller},\ and\ \citenamefont
  {Poulton}}]{WolffTutorial}%
  \BibitemOpen
  \bibfield  {author} {\bibinfo {author} {\bibfnamefont {C.}~\bibnamefont
  {Wolff}}, \bibinfo {author} {\bibfnamefont {M.~J.~A.}\ \bibnamefont {Smith}},
  \bibinfo {author} {\bibfnamefont {B.}~\bibnamefont {Stiller}},\ and\ \bibinfo
  {author} {\bibfnamefont {C.~G.}\ \bibnamefont {Poulton}},\ }\bibfield
  {title} {\bibinfo {title} {Brillouin scattering---theory and experiment:
  Tutorial},\ }\href {https://doi.org/10.1364/JOSAB.416747} {\bibfield
  {journal} {\bibinfo  {journal} {JOSA B}\ }\textbf {\bibinfo {volume} {38}},\
  \bibinfo {pages} {1243} (\bibinfo {year} {2021})}\BibitemShut {NoStop}%
\bibitem [{\citenamefont {Clark}\ \emph {et~al.}(2017)\citenamefont {Clark},
  \citenamefont {Lecocq}, \citenamefont {Simmonds}, \citenamefont {Aumentado},\
  and\ \citenamefont {Teufel}}]{clark2017sideband}%
  \BibitemOpen
  \bibfield  {author} {\bibinfo {author} {\bibfnamefont {J.~B.}\ \bibnamefont
  {Clark}}, \bibinfo {author} {\bibfnamefont {F.}~\bibnamefont {Lecocq}},
  \bibinfo {author} {\bibfnamefont {R.~W.}\ \bibnamefont {Simmonds}}, \bibinfo
  {author} {\bibfnamefont {J.}~\bibnamefont {Aumentado}},\ and\ \bibinfo
  {author} {\bibfnamefont {J.~D.}\ \bibnamefont {Teufel}},\ }\bibfield  {title}
  {\bibinfo {title} {Sideband cooling beyond the quantum backaction limit with
  squeezed light},\ }\href {https://doi.org/10.1038/nature20604} {\bibfield
  {journal} {\bibinfo  {journal} {Nature}\ }\textbf {\bibinfo {volume} {541}},\
  \bibinfo {pages} {191} (\bibinfo {year} {2017})}\BibitemShut {NoStop}%
\bibitem [{\citenamefont {{Wilson-Rae}}\ \emph {et~al.}(2007)\citenamefont
  {{Wilson-Rae}}, \citenamefont {Nooshi}, \citenamefont {Zwerger},\ and\
  \citenamefont {Kippenberg}}]{wilson2007theory}%
  \BibitemOpen
  \bibfield  {author} {\bibinfo {author} {\bibfnamefont {I.}~\bibnamefont
  {{Wilson-Rae}}}, \bibinfo {author} {\bibfnamefont {N.}~\bibnamefont
  {Nooshi}}, \bibinfo {author} {\bibfnamefont {W.}~\bibnamefont {Zwerger}},\
  and\ \bibinfo {author} {\bibfnamefont {T.~J.}\ \bibnamefont {Kippenberg}},\
  }\bibfield  {title} {\bibinfo {title} {Theory of {{Ground State Cooling}} of
  a {{Mechanical Oscillator Using Dynamical Backaction}}},\ }\href
  {https://doi.org/10.1103/PhysRevLett.99.093901} {\bibfield  {journal}
  {\bibinfo  {journal} {Physical Review Letters}\ }\textbf {\bibinfo {volume}
  {99}},\ \bibinfo {pages} {093901} (\bibinfo {year} {2007})}\BibitemShut
  {NoStop}%
\bibitem [{\citenamefont {Zhang}\ \emph {et~al.}(2023)\citenamefont {Zhang},
  \citenamefont {Zhu}, \citenamefont {Wolff},\ and\ \citenamefont
  {Stiller}}]{ZhangCohCtrl}%
  \BibitemOpen
  \bibfield  {author} {\bibinfo {author} {\bibfnamefont {J.}~\bibnamefont
  {Zhang}}, \bibinfo {author} {\bibfnamefont {C.}~\bibnamefont {Zhu}}, \bibinfo
  {author} {\bibfnamefont {C.}~\bibnamefont {Wolff}},\ and\ \bibinfo {author}
  {\bibfnamefont {B.}~\bibnamefont {Stiller}},\ }\bibfield  {title} {\bibinfo
  {title} {Quantum coherent control in pulsed waveguide optomechanics},\ }\href
  {https://doi.org/10.1103/PhysRevResearch.5.013010} {\bibfield  {journal}
  {\bibinfo  {journal} {Physical Review Research}\ }\textbf {\bibinfo {volume}
  {5}},\ \bibinfo {pages} {013010} (\bibinfo {year} {2023})}\BibitemShut
  {NoStop}%
\bibitem [{\citenamefont {Stiller}\ \emph {et~al.}(2019)\citenamefont
  {Stiller}, \citenamefont {Merklein}, \citenamefont {Vu}, \citenamefont {Ma},
  \citenamefont {Madden}, \citenamefont {Poulton},\ and\ \citenamefont
  {Eggleton}}]{stiller2019cross}%
  \BibitemOpen
  \bibfield  {author} {\bibinfo {author} {\bibfnamefont {B.}~\bibnamefont
  {Stiller}}, \bibinfo {author} {\bibfnamefont {M.}~\bibnamefont {Merklein}},
  \bibinfo {author} {\bibfnamefont {K.}~\bibnamefont {Vu}}, \bibinfo {author}
  {\bibfnamefont {P.}~\bibnamefont {Ma}}, \bibinfo {author} {\bibfnamefont
  {S.~J.}\ \bibnamefont {Madden}}, \bibinfo {author} {\bibfnamefont {C.~G.}\
  \bibnamefont {Poulton}},\ and\ \bibinfo {author} {\bibfnamefont {B.~J.}\
  \bibnamefont {Eggleton}},\ }\bibfield  {title} {\bibinfo {title} {Cross
  talk-free coherent multi-wavelength {{Brillouin}} interaction},\ }\href
  {https://doi.org/10.1063/1.5087180} {\bibfield  {journal} {\bibinfo
  {journal} {APL Photonics}\ }\textbf {\bibinfo {volume} {4}},\ \bibinfo
  {pages} {040802} (\bibinfo {year} {2019})}\BibitemShut {NoStop}%
\bibitem [{\citenamefont {Cattiaux}\ \emph {et~al.}(2021)\citenamefont
  {Cattiaux}, \citenamefont {Golokolenov}, \citenamefont {Kumar}, \citenamefont
  {Sillanp{\"a}{\"a}}, \citenamefont {{Mercier de L{\'e}pinay}}, \citenamefont
  {Gazizulin}, \citenamefont {Zhou}, \citenamefont {Armour}, \citenamefont
  {Bourgeois}, \citenamefont {Fefferman},\ and\ \citenamefont
  {Collin}}]{cattiaux2021macroscopic}%
  \BibitemOpen
  \bibfield  {author} {\bibinfo {author} {\bibfnamefont {D.}~\bibnamefont
  {Cattiaux}}, \bibinfo {author} {\bibfnamefont {I.}~\bibnamefont
  {Golokolenov}}, \bibinfo {author} {\bibfnamefont {S.}~\bibnamefont {Kumar}},
  \bibinfo {author} {\bibfnamefont {M.}~\bibnamefont {Sillanp{\"a}{\"a}}},
  \bibinfo {author} {\bibfnamefont {L.}~\bibnamefont {{Mercier de
  L{\'e}pinay}}}, \bibinfo {author} {\bibfnamefont {R.~R.}\ \bibnamefont
  {Gazizulin}}, \bibinfo {author} {\bibfnamefont {X.}~\bibnamefont {Zhou}},
  \bibinfo {author} {\bibfnamefont {A.~D.}\ \bibnamefont {Armour}}, \bibinfo
  {author} {\bibfnamefont {O.}~\bibnamefont {Bourgeois}}, \bibinfo {author}
  {\bibfnamefont {A.}~\bibnamefont {Fefferman}},\ and\ \bibinfo {author}
  {\bibfnamefont {E.}~\bibnamefont {Collin}},\ }\bibfield  {title} {\bibinfo
  {title} {A macroscopic object passively cooled into its quantum ground state
  of motion beyond single-mode cooling},\ }\href
  {https://doi.org/10.1038/s41467-021-26457-8} {\bibfield  {journal} {\bibinfo
  {journal} {Nature Communications}\ }\textbf {\bibinfo {volume} {12}},\
  \bibinfo {pages} {6182} (\bibinfo {year} {2021})}\BibitemShut {NoStop}%
\bibitem [{\citenamefont {Schrinski}\ \emph {et~al.}(2023)\citenamefont
  {Schrinski}, \citenamefont {Yang}, \citenamefont {{von L{\"u}pke}},
  \citenamefont {Bild}, \citenamefont {Chu}, \citenamefont {Hornberger},
  \citenamefont {Nimmrichter},\ and\ \citenamefont {Fadel}}]{MacroscopQuantum}%
  \BibitemOpen
  \bibfield  {author} {\bibinfo {author} {\bibfnamefont {B.}~\bibnamefont
  {Schrinski}}, \bibinfo {author} {\bibfnamefont {Y.}~\bibnamefont {Yang}},
  \bibinfo {author} {\bibfnamefont {U.}~\bibnamefont {{von L{\"u}pke}}},
  \bibinfo {author} {\bibfnamefont {M.}~\bibnamefont {Bild}}, \bibinfo {author}
  {\bibfnamefont {Y.}~\bibnamefont {Chu}}, \bibinfo {author} {\bibfnamefont
  {K.}~\bibnamefont {Hornberger}}, \bibinfo {author} {\bibfnamefont
  {S.}~\bibnamefont {Nimmrichter}},\ and\ \bibinfo {author} {\bibfnamefont
  {M.}~\bibnamefont {Fadel}},\ }\bibfield  {title} {\bibinfo {title}
  {Macroscopic {{Quantum Test}} with {{Bulk Acoustic Wave Resonators}}},\
  }\href {https://doi.org/10.1103/PhysRevLett.130.133604} {\bibfield  {journal}
  {\bibinfo  {journal} {Physical Review Letters}\ }\textbf {\bibinfo {volume}
  {130}},\ \bibinfo {pages} {133604} (\bibinfo {year} {2023})}\BibitemShut
  {NoStop}%
\bibitem [{\citenamefont {Merklein}\ \emph {et~al.}(2017)\citenamefont
  {Merklein}, \citenamefont {Stiller}, \citenamefont {Vu}, \citenamefont
  {Madden},\ and\ \citenamefont {Eggleton}}]{merklein2017chip}%
  \BibitemOpen
  \bibfield  {author} {\bibinfo {author} {\bibfnamefont {M.}~\bibnamefont
  {Merklein}}, \bibinfo {author} {\bibfnamefont {B.}~\bibnamefont {Stiller}},
  \bibinfo {author} {\bibfnamefont {K.}~\bibnamefont {Vu}}, \bibinfo {author}
  {\bibfnamefont {S.~J.}\ \bibnamefont {Madden}},\ and\ \bibinfo {author}
  {\bibfnamefont {B.~J.}\ \bibnamefont {Eggleton}},\ }\bibfield  {title}
  {\bibinfo {title} {A chip-integrated coherent photonic-phononic memory},\
  }\href {https://doi.org/10.1038/s41467-017-00717-y} {\bibfield  {journal}
  {\bibinfo  {journal} {Nature Communications}\ }\textbf {\bibinfo {volume}
  {8}},\ \bibinfo {pages} {574} (\bibinfo {year} {2017})}\BibitemShut {NoStop}%
\bibitem [{\citenamefont {Otterstrom}\ \emph {et~al.}(2018)\citenamefont
  {Otterstrom}, \citenamefont {Behunin}, \citenamefont {Kittlaus},\ and\
  \citenamefont {Rakich}}]{Otterstrom}%
  \BibitemOpen
  \bibfield  {author} {\bibinfo {author} {\bibfnamefont {N.~T.}\ \bibnamefont
  {Otterstrom}}, \bibinfo {author} {\bibfnamefont {R.~O.}\ \bibnamefont
  {Behunin}}, \bibinfo {author} {\bibfnamefont {E.~A.}\ \bibnamefont
  {Kittlaus}},\ and\ \bibinfo {author} {\bibfnamefont {P.~T.}\ \bibnamefont
  {Rakich}},\ }\bibfield  {title} {\bibinfo {title} {Optomechanical {{Cooling}}
  in a {{Continuous System}}},\ }\href
  {https://doi.org/10.1103/PhysRevX.8.041034} {\bibfield  {journal} {\bibinfo
  {journal} {Physical Review X}\ }\textbf {\bibinfo {volume} {8}},\ \bibinfo
  {pages} {041034} (\bibinfo {year} {2018})}\BibitemShut {NoStop}%
\bibitem [{\citenamefont {Johnson}\ \emph {et~al.}(2023)\citenamefont
  {Johnson}, \citenamefont {Haverkamp}, \citenamefont {Ou}, \citenamefont
  {Kieu}, \citenamefont {Otterstrom}, \citenamefont {Rakich},\ and\
  \citenamefont {Behunin}}]{JohnsonCooling}%
  \BibitemOpen
  \bibfield  {author} {\bibinfo {author} {\bibfnamefont {J.~N.}\ \bibnamefont
  {Johnson}}, \bibinfo {author} {\bibfnamefont {D.~R.}\ \bibnamefont
  {Haverkamp}}, \bibinfo {author} {\bibfnamefont {Y.-H.}\ \bibnamefont {Ou}},
  \bibinfo {author} {\bibfnamefont {K.}~\bibnamefont {Kieu}}, \bibinfo {author}
  {\bibfnamefont {N.~T.}\ \bibnamefont {Otterstrom}}, \bibinfo {author}
  {\bibfnamefont {P.~T.}\ \bibnamefont {Rakich}},\ and\ \bibinfo {author}
  {\bibfnamefont {R.~O.}\ \bibnamefont {Behunin}},\ }\bibfield  {title}
  {\bibinfo {title} {Laser {{Cooling}} of {{Traveling-Wave Phonons}} in an
  {{Optical Fiber}}},\ }\href
  {https://doi.org/10.1103/PhysRevApplied.20.034047} {\bibfield  {journal}
  {\bibinfo  {journal} {Physical Review Applied}\ }\textbf {\bibinfo {volume}
  {20}},\ \bibinfo {pages} {034047} (\bibinfo {year} {2023})}\BibitemShut
  {NoStop}%
\bibitem [{\citenamefont {Zhu}\ \emph {et~al.}(2024)\citenamefont {Zhu},
  \citenamefont {Genes},\ and\ \citenamefont {Stiller}}]{zhu2024optoacoustic}%
  \BibitemOpen
  \bibfield  {author} {\bibinfo {author} {\bibfnamefont {C.}~\bibnamefont
  {Zhu}}, \bibinfo {author} {\bibfnamefont {C.}~\bibnamefont {Genes}},\ and\
  \bibinfo {author} {\bibfnamefont {B.}~\bibnamefont {Stiller}},\ }\bibfield
  {title} {\bibinfo {title} {Optoacoustic {{Entanglement}} in a {{Continuous
  Brillouin-Active Solid State System}}},\ }\href
  {https://doi.org/10.1103/PhysRevLett.133.203602} {\bibfield  {journal}
  {\bibinfo  {journal} {Physical Review Letters}\ }\textbf {\bibinfo {volume}
  {133}},\ \bibinfo {pages} {203602} (\bibinfo {year} {2024})}\BibitemShut
  {NoStop}%
\bibitem [{\citenamefont {Tow}\ \emph {et~al.}(2012)\citenamefont {Tow},
  \citenamefont {L{\'e}guillon}, \citenamefont {Fresnel}, \citenamefont
  {Besnard}, \citenamefont {Brilland}, \citenamefont {M{\'e}chin},
  \citenamefont {Tr{\'e}goat}, \citenamefont {Troles},\ and\ \citenamefont
  {Toupin}}]{tow2012linewidth}%
  \BibitemOpen
  \bibfield  {author} {\bibinfo {author} {\bibfnamefont {K.~H.}\ \bibnamefont
  {Tow}}, \bibinfo {author} {\bibfnamefont {Y.}~\bibnamefont {L{\'e}guillon}},
  \bibinfo {author} {\bibfnamefont {S.}~\bibnamefont {Fresnel}}, \bibinfo
  {author} {\bibfnamefont {P.}~\bibnamefont {Besnard}}, \bibinfo {author}
  {\bibfnamefont {L.}~\bibnamefont {Brilland}}, \bibinfo {author}
  {\bibfnamefont {D.}~\bibnamefont {M{\'e}chin}}, \bibinfo {author}
  {\bibfnamefont {D.}~\bibnamefont {Tr{\'e}goat}}, \bibinfo {author}
  {\bibfnamefont {J.}~\bibnamefont {Troles}},\ and\ \bibinfo {author}
  {\bibfnamefont {P.}~\bibnamefont {Toupin}},\ }\bibfield  {title} {\bibinfo
  {title} {Linewidth-narrowing and intensity noise reduction of the 2nd order
  {{Stokes}} component of a low threshold {{Brillouin}} laser made of
  {{Ge}}{\textsubscript{10}}{{As}}{\textsubscript{22}}{{Se}}{\textsubscript{68}}
  chalcogenide fiber},\ }\href {https://doi.org/10.1364/OE.20.00B104}
  {\bibfield  {journal} {\bibinfo  {journal} {Optics Express}\ }\textbf
  {\bibinfo {volume} {20}},\ \bibinfo {pages} {B104} (\bibinfo {year}
  {2012})}\BibitemShut {NoStop}%
\bibitem [{\citenamefont {Tow}(2013)}]{tow2013laser}%
  \BibitemOpen
  \bibfield  {author} {\bibinfo {author} {\bibfnamefont {K.~H.}\ \bibnamefont
  {Tow}},\ }\emph {\bibinfo {title} {{Laser Brillouin {\`a} fibre
  microstructur{\'e}e en verre de chaleog{\'e}nure}}},\ \href@noop {} {Ph.D.
  thesis},\ \bibinfo  {school} {Universit{\'e} de Rennes} (\bibinfo {year}
  {2013})\BibitemShut {NoStop}%
\bibitem [{\citenamefont {{Cryer-Jenkins}}\ \emph {et~al.}(2025)\citenamefont
  {{Cryer-Jenkins}}, \citenamefont {Leung}, \citenamefont {Rathee},
  \citenamefont {Tan}, \citenamefont {Major},\ and\ \citenamefont
  {Vanner}}]{cryer2025brillouin}%
  \BibitemOpen
  \bibfield  {author} {\bibinfo {author} {\bibfnamefont {E.~A.}\ \bibnamefont
  {{Cryer-Jenkins}}}, \bibinfo {author} {\bibfnamefont {A.~C.}\ \bibnamefont
  {Leung}}, \bibinfo {author} {\bibfnamefont {H.}~\bibnamefont {Rathee}},
  \bibinfo {author} {\bibfnamefont {A.~K.~C.}\ \bibnamefont {Tan}}, \bibinfo
  {author} {\bibfnamefont {K.~D.}\ \bibnamefont {Major}},\ and\ \bibinfo
  {author} {\bibfnamefont {M.~R.}\ \bibnamefont {Vanner}},\ }\bibfield  {title}
  {\bibinfo {title} {Brillouin--{{Mandelstam}} scattering in telecommunications
  optical fiber at millikelvin temperatures},\ }\href
  {https://doi.org/10.1063/5.0241253} {\bibfield  {journal} {\bibinfo
  {journal} {APL Photonics}\ }\textbf {\bibinfo {volume} {10}},\ \bibinfo
  {pages} {010805} (\bibinfo {year} {2025})}\BibitemShut {NoStop}%
\bibitem [{\citenamefont {Floch}\ \emph {et~al.}(2001)\citenamefont {Floch},
  \citenamefont {Riou},\ and\ \citenamefont {Cambon}}]{le2001experimental}%
  \BibitemOpen
  \bibfield  {author} {\bibinfo {author} {\bibfnamefont {S.~L.}\ \bibnamefont
  {Floch}}, \bibinfo {author} {\bibfnamefont {F.}~\bibnamefont {Riou}},\ and\
  \bibinfo {author} {\bibfnamefont {P.}~\bibnamefont {Cambon}},\ }\bibfield
  {title} {\bibinfo {title} {Experimental and theoretical study of the
  {{Brillouin}} linewidth and frequency at low temperature in standard
  single-mode optical fibres},\ }\href
  {https://doi.org/10.1088/1464-4258/3/3/102} {\bibfield  {journal} {\bibinfo
  {journal} {Journal of Optics A: Pure and Applied Optics}\ }\textbf {\bibinfo
  {volume} {3}},\ \bibinfo {pages} {L12} (\bibinfo {year} {2001})}\BibitemShut
  {NoStop}%
\bibitem [{\citenamefont {Le~Floch}\ and\ \citenamefont
  {Cambon}(2003)}]{le2003study}%
  \BibitemOpen
  \bibfield  {author} {\bibinfo {author} {\bibfnamefont {S.}~\bibnamefont
  {Le~Floch}}\ and\ \bibinfo {author} {\bibfnamefont {P.}~\bibnamefont
  {Cambon}},\ }\bibfield  {title} {\bibinfo {title} {Study of {{Brillouin}}
  gain spectrum in standard single-mode optical fiber at low temperatures
  (1.4--370 {{K}}) and high hydrostatic pressures (1--250 bars)},\ }\href
  {https://doi.org/10.1016/S0030-4018(03)01296-3} {\bibfield  {journal}
  {\bibinfo  {journal} {Optics Communications}\ }\textbf {\bibinfo {volume}
  {219}},\ \bibinfo {pages} {395} (\bibinfo {year} {2003})}\BibitemShut
  {NoStop}%
\bibitem [{\citenamefont {Wang}\ \emph {et~al.}(2020)\citenamefont {Wang},
  \citenamefont {Gao}, \citenamefont {Baker}, \citenamefont {Wang},
  \citenamefont {Chen},\ and\ \citenamefont {Bao}}]{wang2020stimulated}%
  \BibitemOpen
  \bibfield  {author} {\bibinfo {author} {\bibfnamefont {H.}~\bibnamefont
  {Wang}}, \bibinfo {author} {\bibfnamefont {S.}~\bibnamefont {Gao}}, \bibinfo
  {author} {\bibfnamefont {C.}~\bibnamefont {Baker}}, \bibinfo {author}
  {\bibfnamefont {Y.}~\bibnamefont {Wang}}, \bibinfo {author} {\bibfnamefont
  {L.}~\bibnamefont {Chen}},\ and\ \bibinfo {author} {\bibfnamefont
  {X.}~\bibnamefont {Bao}},\ }\bibfield  {title} {\bibinfo {title} {Stimulated
  {{Brillouin}} scattering in a tapered dual-core
  {{As}}{\textsubscript{2}}{{Se}}{\textsubscript{3}}-{{PMMA}} fiber for
  simultaneous temperature and strain sensing},\ }\href
  {https://doi.org/10.1364/OL.391734} {\bibfield  {journal} {\bibinfo
  {journal} {Optics Letters}\ }\textbf {\bibinfo {volume} {45}},\ \bibinfo
  {pages} {3301} (\bibinfo {year} {2020})}\BibitemShut {NoStop}%
\bibitem [{\citenamefont {Mart{\'i}nez}\ \emph {et~al.}(2025)\citenamefont
  {Mart{\'i}nez}, \citenamefont {Zhu},\ and\ \citenamefont
  {Stiller}}]{martínez2025cavitylessbrillouinstrongcoupling}%
  \BibitemOpen
  \bibfield  {author} {\bibinfo {author} {\bibfnamefont {L.~B.}\ \bibnamefont
  {Mart{\'i}nez}}, \bibinfo {author} {\bibfnamefont {C.}~\bibnamefont {Zhu}},\
  and\ \bibinfo {author} {\bibfnamefont {B.}~\bibnamefont {Stiller}},\ }\href
  {https://doi.org/10.48550/arXiv.2507.08673} {\bibinfo {title} {Cavity-less
  {{Brillouin}} strong coupling in a solid-state continuous system}} (\bibinfo
  {year} {2025}),\ \Eprint {https://arxiv.org/abs/2507.08673} {arXiv:2507.08673
  [physics]} \BibitemShut {NoStop}%
\bibitem [{\citenamefont {Chou}\ \emph {et~al.}(2025)\citenamefont {Chou},
  \citenamefont {Qiao}, \citenamefont {Yan}, \citenamefont {Andersson},
  \citenamefont {Conner}, \citenamefont {Grebel}, \citenamefont {Joshi},
  \citenamefont {Miller}, \citenamefont {Povey}, \citenamefont {Wu},\ and\
  \citenamefont {Cleland}}]{chou2025phononEntanglement}%
  \BibitemOpen
  \bibfield  {author} {\bibinfo {author} {\bibfnamefont {M.-H.}\ \bibnamefont
  {Chou}}, \bibinfo {author} {\bibfnamefont {H.}~\bibnamefont {Qiao}}, \bibinfo
  {author} {\bibfnamefont {H.}~\bibnamefont {Yan}}, \bibinfo {author}
  {\bibfnamefont {G.}~\bibnamefont {Andersson}}, \bibinfo {author}
  {\bibfnamefont {C.~R.}\ \bibnamefont {Conner}}, \bibinfo {author}
  {\bibfnamefont {J.}~\bibnamefont {Grebel}}, \bibinfo {author} {\bibfnamefont
  {Y.~J.}\ \bibnamefont {Joshi}}, \bibinfo {author} {\bibfnamefont {J.~M.}\
  \bibnamefont {Miller}}, \bibinfo {author} {\bibfnamefont {R.~G.}\
  \bibnamefont {Povey}}, \bibinfo {author} {\bibfnamefont {X.}~\bibnamefont
  {Wu}},\ and\ \bibinfo {author} {\bibfnamefont {A.~N.}\ \bibnamefont
  {Cleland}},\ }\bibfield  {title} {\bibinfo {title} {Deterministic
  multi-phonon entanglement between two mechanical resonators on separate
  substrates},\ }\href {https://doi.org/10.1038/s41467-025-56454-0} {\bibfield
  {journal} {\bibinfo  {journal} {Nature Communications}\ }\textbf {\bibinfo
  {volume} {16}},\ \bibinfo {pages} {1450} (\bibinfo {year}
  {2025})}\BibitemShut {NoStop}%
\bibitem [{\citenamefont {Mirhosseini}\ \emph {et~al.}(2020)\citenamefont
  {Mirhosseini}, \citenamefont {Sipahigil}, \citenamefont {Kalaee},\ and\
  \citenamefont {Painter}}]{mirhosseini2020superconducting}%
  \BibitemOpen
  \bibfield  {author} {\bibinfo {author} {\bibfnamefont {M.}~\bibnamefont
  {Mirhosseini}}, \bibinfo {author} {\bibfnamefont {A.}~\bibnamefont
  {Sipahigil}}, \bibinfo {author} {\bibfnamefont {M.}~\bibnamefont {Kalaee}},\
  and\ \bibinfo {author} {\bibfnamefont {O.}~\bibnamefont {Painter}},\
  }\bibfield  {title} {\bibinfo {title} {Superconducting qubit to optical
  photon transduction},\ }\href {https://doi.org/10.1038/s41586-020-3038-6}
  {\bibfield  {journal} {\bibinfo  {journal} {Nature}\ }\textbf {\bibinfo
  {volume} {588}},\ \bibinfo {pages} {599} (\bibinfo {year}
  {2020})}\BibitemShut {NoStop}%
\bibitem [{\citenamefont {Guo}\ \emph {et~al.}(2019)\citenamefont {Guo},
  \citenamefont {Feng}, \citenamefont {Yang}, \citenamefont {Yu}, \citenamefont
  {Chen}, \citenamefont {Yuan},\ and\ \citenamefont
  {Zhang}}]{guo2019highQuantMemory}%
  \BibitemOpen
  \bibfield  {author} {\bibinfo {author} {\bibfnamefont {J.}~\bibnamefont
  {Guo}}, \bibinfo {author} {\bibfnamefont {X.}~\bibnamefont {Feng}}, \bibinfo
  {author} {\bibfnamefont {P.}~\bibnamefont {Yang}}, \bibinfo {author}
  {\bibfnamefont {Z.}~\bibnamefont {Yu}}, \bibinfo {author} {\bibfnamefont
  {L.~Q.}\ \bibnamefont {Chen}}, \bibinfo {author} {\bibfnamefont {C.-H.}\
  \bibnamefont {Yuan}},\ and\ \bibinfo {author} {\bibfnamefont
  {W.}~\bibnamefont {Zhang}},\ }\bibfield  {title} {\bibinfo {title}
  {High-performance {{Raman}} quantum memory with optimal control in room
  temperature atoms},\ }\href {https://doi.org/10.1038/s41467-018-08118-5}
  {\bibfield  {journal} {\bibinfo  {journal} {Nature Communications}\ }\textbf
  {\bibinfo {volume} {10}},\ \bibinfo {pages} {148} (\bibinfo {year}
  {2019})}\BibitemShut {NoStop}%
\end{thebibliography}

%

\end{document}